\documentclass[
 aip,
 amsmath,amssymb,
reprint,
twocolumn
floatfix
]{revtex4-2}

\usepackage{graphicx}
\usepackage{import}
\usepackage{dcolumn}
\usepackage{bm}
\usepackage{transparent}
\usepackage[utf8]{inputenc}
\usepackage[T1]{fontenc}
\usepackage{mathptmx}
\usepackage{etoolbox}
\usepackage{siunitx}
\usepackage{xcolor}

\makeatletter
\def\@email#1#2{%
 \endgroup
 \patchcmd{\titleblock@produce}
  {\frontmatter@RRAPformat}
  {\frontmatter@RRAPformat{\produce@RRAP{*#1\href{mailto:#2}{#2}}}\frontmatter@RRAPformat}
  {}{}
}%
\makeatother

\begin{document}

\preprint{AIP/123-QED}

\title[]{Magnetic bound states embedded in tantalum superconducting thin films}

\author{Soroush~Arabi}

\altaffiliation{ Correspondending author: soroush.arabi@kit.edu}
\affiliation{\mbox{Institute for Quantum Materials and Technologies, Karlsruhe Institute of Technology, 76131 Karlsruhe, Germany}}
\affiliation{\mbox{Physikalisches Institut, Karlsruhe Institute of Technology, 76131 Karlsruhe, Germany}}

\author{Qili~Li}%

\affiliation{\mbox{Physikalisches Institut, Karlsruhe Institute of Technology, 76131 Karlsruhe, Germany}}

\author{Ritika~Dhundhwal}

\affiliation{\mbox{Institute for Quantum Materials and Technologies, Karlsruhe Institute of Technology, 76131 Karlsruhe, Germany}}

\author{Dirk~Fuchs}

\affiliation{\mbox{Institute for Quantum Materials and Technologies, Karlsruhe Institute of Technology, 76131 Karlsruhe, Germany}}

\author{Thomas~Reisinger}

\affiliation{\mbox{Institute for Quantum Materials and Technologies, Karlsruhe Institute of Technology, 76131 Karlsruhe, Germany}}
\author{Ioan~M.~Pop}

\affiliation{\mbox{Institute for Quantum Materials and Technologies, Karlsruhe Institute of Technology, 76131 Karlsruhe, Germany}}
\affiliation{\mbox{Physikalisches Institut, Karlsruhe Institute of Technology, 76131 Karlsruhe, Germany}}
\affiliation{\mbox{Physics Institute 1, Stuttgart University, 70569 Stuttgart, Germany}}
\author{Wulf~Wulfhekel}

\affiliation{\mbox{Institute for Quantum Materials and Technologies, Karlsruhe Institute of Technology, 76131 Karlsruhe, Germany}}
\affiliation{\mbox{Physikalisches Institut, Karlsruhe Institute of Technology, 76131 Karlsruhe, Germany}}

\begin{abstract}
In the fabrication of superconducting devices, both \textit{in situ} and \textit{ex situ} processes are utilized, making the removal of unwanted oxide layers and impurities under vacuum conditions crucial. Oxygen descumming and argon milling are standard \textit{in situ} cleaning methods employed for device preparation. We investigated the impact of these techniques on tantalum superconducting thin films using scanning tunneling microscopy at millikelvin temperatures. We demonstrate that these cleaning methods inadvertently introduce magnetic bound states within the superconducting gap of tantalum, likely by oxygen impurities. These bound states can be detrimental to superconducting qubit devices, as they add to dephasing and energy relaxation.
\end{abstract}

\maketitle

\begin{figure}
    \centering
    \def\svgwidth{\columnwidth} 
     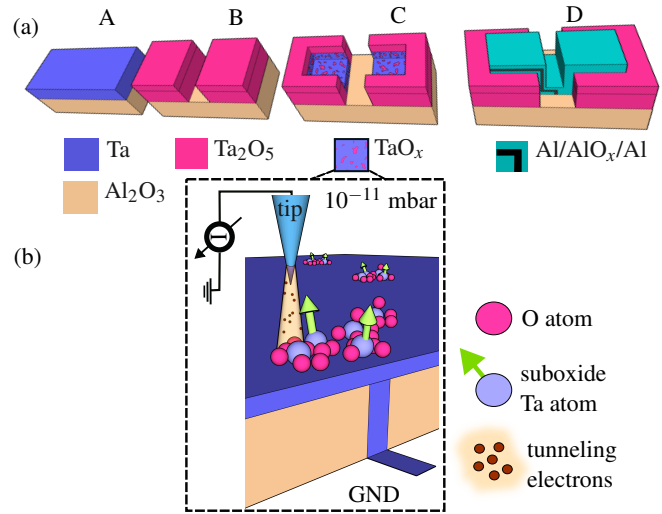
    \caption{Schematic overview of the main steps in tantalum qubit metallization and the principle of STM measurements. (a) The metallization process begins with the Al$_{2}$O$_{3}$ substrate (sand brown) undergoing solvent cleaning. Next, tantalum (majorelle blue) is sputtered onto the substrate using DC magnetron sputtering at 600°C in UHV (A). In (B), the substrate exits UHV for patterning and additional cleaning, during which a layer of Ta$_{2}$O$_{5}$ (red pink) forms. The sample is then returned to UHV for a cleaning protocol that combines O$_{2}$ plasma treatment and argon milling (C). Argon milling does not fully remove the native oxides, leaving non-stoichiometric TaO$_x$ traces (blue-red mix) on or near the surface. The final step (D) involves the deposition of aluminum (cyan), aluminum oxide (black), and aluminum to form Josephson junctions. \textbf{(b)} Scanning tunneling microscopy (STM) is performed on the TaO$_x$ traces using a superconducting Ta tip at 45 mK. The tantalum sample is connected to the ground (GND) via a side-contact.}
    \label{fig:Figure1}
\end{figure}

The quest for quantum computing platforms capable of realizing reliable and scalable qubit devices has spurred intensive research efforts in the last decades \cite{Blais2021, Siddiqi2021}. A key strategy driving progress in this area is the synergistic integration of diverse materials, leveraging their distinct advantages \cite{Leon2021, Liu2019}. One notable example is the tantalum transmon qubit \cite{Place2021}, which incorporates tantalum for the capacitive elements due to its well-behaved surface oxide \cite{McLellan2023} and compatibility with harsh cleaning methods, while aluminum is utilized for the Josephson junction, thanks to its stable and controllable thermal oxide. However, the fabrication process presents a significant challenge: it involves two separate lithography steps, which expose the tantalum surface to air, introducing surface contamination and oxidation (see Figure \ref{fig:Figure1}a).

Maintaining a pristine tantalum contact surface and a clean tantalum-aluminum interface is essential to ensure a high-quality circuit. This requires the effective removal of e-beam residues from lithography and the oxide layer that forms after breaking the vacuum. Traditionally, to connect the tantalum and aluminum layers, a step of oxygen descumming (plasma/ash) followed by argon milling of the tantalum layer is used. Oxygen plasma effectively volatilizes organic materials, such as residual e-beam resist molecules, aiding in their removal but simultaneously promoting additional surface oxidation. Argon milling is then employed to remove both the oxide layers and any remaining volatile organic molecules, with the goal of achieving a pristine contact surface for subsequent aluminum deposition. Our study, however, reveals an unanticipated consequence of this cleaning procedure: the creation of magnetic bound states at the interface by argon milling, which can act as a source of spurious two-level systems and pose further challenges to qubit performance.

When tantalum is exposed to air, its surface becomes covered by a fully oxidized Ta$_{2}$O$_{5}$ layer, renowned for its stability and insulating properties \cite{McLellan2023, Devine1996, CHANELIERE1998}. Beneath this oxide layer, suboxide species such as Ta$^{3+}$ and Ta$^{1+}$ exist in minor quantities within an amorphous transition region between the oxide and the metallic bulk \cite{McLellan2023}. The removal of the surface oxide by Ar$^+$ milling involves a stochastic scattering process, which can generate vacancies, interstitials, and impurities within the tantalum film. In particular, intermixed oxygen can induce localized states in the thin film with charge and spin degrees of freedom \cite{Ivanov2011, Shu2022, Yang2020, Ramprasad2003, Perez2019}. While aiming for the complete removal of the oxide layer, intensive Ar$^+$ milling may introduce numerous defects in both the film and the substrate, whereas moderate milling typically leaves traces of these localized states.
\begin{figure*}
    \centering
     \def\svgwidth{2\columnwidth} 
     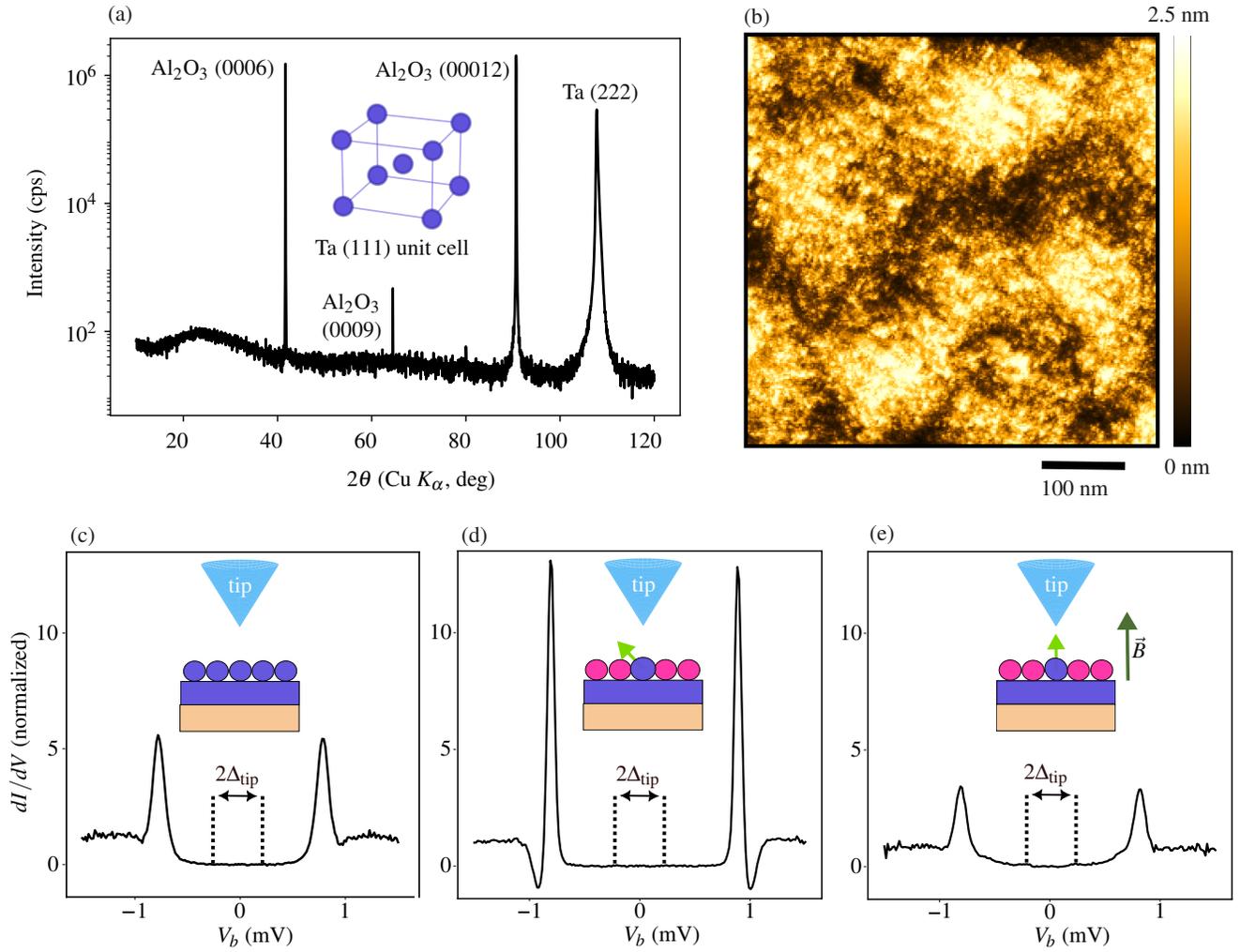
    \caption{XRD and STM characterization of an $\alpha$(111) tantalum film. (b) XRD spectrum of the tantalum film grown on c-plane Al$_{2}$O$_{3}$. The diffraction pattern reveals a single Ta peak, labeled as $\alpha$(222), indicating the $\alpha$ phase growing in the (111) direction. The $\alpha$(111) peak is suppressed in diffractograms of BCC crystals (unit cell shown in the inset). A narrow peak at $\ang{65}$ corresponds to diffraction from residual K$_{\beta}$ X-rays originating from the sapphire substrate. (b) STM topography acquired at a bias voltage $V_{b} = 100$ mV and a current setpoint $I_{p} = 90$ pA. (c) Representative $dI/dV$ spectrum displaying the pristine superconducting gap of the combined Ta sample and tip. (d) $dI/dV$ spectrum measured on a suboxide region, showing a pair of effective $S = 1/2$ YSR states with a binding energy of $597.5$ $\mu$eV after Ar$^{+}$ milling. (e) $dI/dV$ spectrum obtained from the same suboxide region as in (d) under an $17$ mT out-of-plane magnetic field. The YSR peaks are suppressed, and the coherence peaks of the sample are reduced compared to (c). Measurements were conducted at $V_{b} = 1.5$ mV, $I_{p} = 100$ pA, and base temperature of $T = 45$ mK using a lock-in technique with a modulation frequency of 3.3 kHz and an AC bias modulation amplitude of $V_{\text{mod}} = 20$ $\mu$V. The spectra incorporate the combined SIS gap for the tantalum film and tip, with $\Delta_{\text{sample}} = 600.3$ $\mu$eV and $\Delta_{\text{tip}} = 208.1$ $\mu$eV.}
    \label{fig:Figure2}
\end{figure*}

Here, we demonstrate that argon milling of oxidized tantalum films induces defects on the surface of the thin film, which, in close proximity to the tantalum superconductor, give rise to localized magnetic bound states, known as Yu-Shiba-Rusinov (YSR) states \cite{Yu1965,Shiba1968,Rusinov1969}. These states represent low-energy excitations within the superconducting energy which can be measured directly by a local probe (see Figure \ref{fig:Figure1}b). The localized magnetic moments exert an exchange interaction $J$ on the Cooper pairs in the superconducting condensate. Depending on the strength of $J$, either the Cooper pairs are broken and the local spin is screened by normal electrons, similar to the Kondo effect (for $\Delta<J$), or the local spin remains unscreened, exerting a small effective field on the Cooper pairs (for $J<\Delta$). The excitation between these two states costs less energy than $\Delta$ and manifests itself as a YSR state within the superconducting gap \cite{Flatt1997,Franke2018}.

The energy of the YSR states is primarily determined by the strength of the exchange coupling, neglecting the Coulomb scattering effect. It is expressed as $\varepsilon = \pm \Delta \frac{1 - \alpha^2}{1 + \alpha^2}$, where $\alpha = JS_{\text{imp}}\pi\rho_{n}$, with $S_{\text{imp}}$ the local moment spin and $\rho_n$ the density of normal electronic states at the Fermi level \cite{Balatsky2006}. In tunneling spectroscopy $dI/dV$, YSR states appear as prominent peaks symmetrically located within the energy gap for $J/\Delta \neq 0$. As $J/\Delta \rightarrow 0$, they shift toward the quasiparticle coherence peaks at $\varepsilon \approx \pm \Delta$, representing an almost free-spin state. The number of peaks increases as the local moment spin increases, from a single pair for spin-$1/2$ to multiple pairs for higher spin states \cite{Ruby2015,Zitko2011}.

We prepared the sample by depositing 200 nm tantalum film on a c-plane sapphire (Al$_{2}$O$_{3}$) substrate at 600 °C using DC magnetron sputtering, resulting in textured film of the $\alpha$ phase oriented predominantly with the epitaxial alignment $(0001)$ Al$_{2}$O$_{3}\,\parallel\,(111)\,$Ta. To determine the crystallographic phase and growth direction of the film, we performed X-ray diffraction (XRD) on the sample. We obtained a $2 \theta$ scan using a monochromatized characteristic X-ray K$_\alpha$ line from a Cu source. The XRD spectrum (Figure \ref{fig:Figure2}a) shows the presence of $\alpha$ phase growing in (111) direction, which is thermodynamically stable and has body-centered cubic structure.

After exposure to air, a native tantalum oxide layer formed on the surface. The sample was then transferred to an ultrahigh vacuum (UHV) chamber for Ar$^{+}$ milling in 15- and 30-minute intervals. The milling process utilized a  differentially pumped extractor-type ion source (SPECS IQE 12/38), delivering an ion beam current density of $3.2$ $\mu \text{A}/\text{cm}^{2}$ at the sample plate with a primary ion energy of $3$ keV. The chamber pressure was maintained at $\sim 2\times 10^{-8}$ mbar throughout the milling procedure. Without breaking the UHV, the sample was subsequently transferred to the scanning tunneling microscope (STM) for surface characterization at $45$ mK using a superconducting tantalum-coated tip  \cite{Balashov2018}. The superconducting tip was prepared by indentation of a tungsten tip on the Ta surface. To enable charge transport in the STM experiments, the samples included a side contact with the ground.

We characterized the surface of the air-exposed tantalum sample after 15 minutes of Ar$^+$ milling. The STM topography (Figure \ref{fig:Figure2}b) reveals a surface with a root-mean-square roughness of only 445.1 pm over an area of 0.5 $(\mu\text{m})^{2}$. Differential conductance ($dI/dV$) measurements on the bare tantalum surface (Figure \ref{fig:Figure2}c) indicate a combined superconducting gap ($2\Delta_{\text{tot.}} = 2\Delta_{\text{sample}} + 2\Delta_{\text{tip}}$) characteristic for tunneling between two superconductors, i.e., in a superconductor-insulator-superconductor (SIS) tunnel junction. To disentangle the contributions of the sample and tip to the SIS spectrum, we performed $dI/dV$ measurements in an out-of-plane magnetic field (see supplementary Figure S1). We extracted zero-field superconducting gaps of $\Delta_{\text{sample,0}} = 600.3$ $\mu$eV and $\Delta_{\text{tip,0}} = 208.1$ $\mu$eV, with corresponding critical fields of $B_{\text{sample,c}} = 103.1$ mT and $B_{\text{tip,c}} = 1.59$ T for sample and tip. The surface gap is slightly smaller than the bulk value of $\Delta_{\text{bulk}} \approx 720$ $\mu$eV for $\alpha$ phase tantalum \cite{Hauser1964, Crowley2023, Schijndel2024}, likely due to small grain size and surface disorder. Furthermore, we frequently observed sharp and prominent single-pair YSR states, positioned asymptotically close to the quasiparticle coherence peaks that dominate the spectrum (Figure \ref{fig:Figure2}d). The binding energy of these states remains close to the coherence peaks of tantalum due to the very weak coupling between local spins and Cooper pairs. When an external magnetic field is applied, even a slight reduction in the superconducting gap drives the bound states out of the gap, leading to their collapse at 17 mT (Figure \ref{fig:Figure2}e). Notably, this modification in the gap structure is accompanied by a significant broadening of the quasiparticle coherence peaks, with their height reduced by approximately 40\% relative to the bare gap (Figure \ref{fig:Figure2}c) in the absence of a magnetic field.
\begin{figure}
    \centering
     \def\svgwidth{1\columnwidth} 
     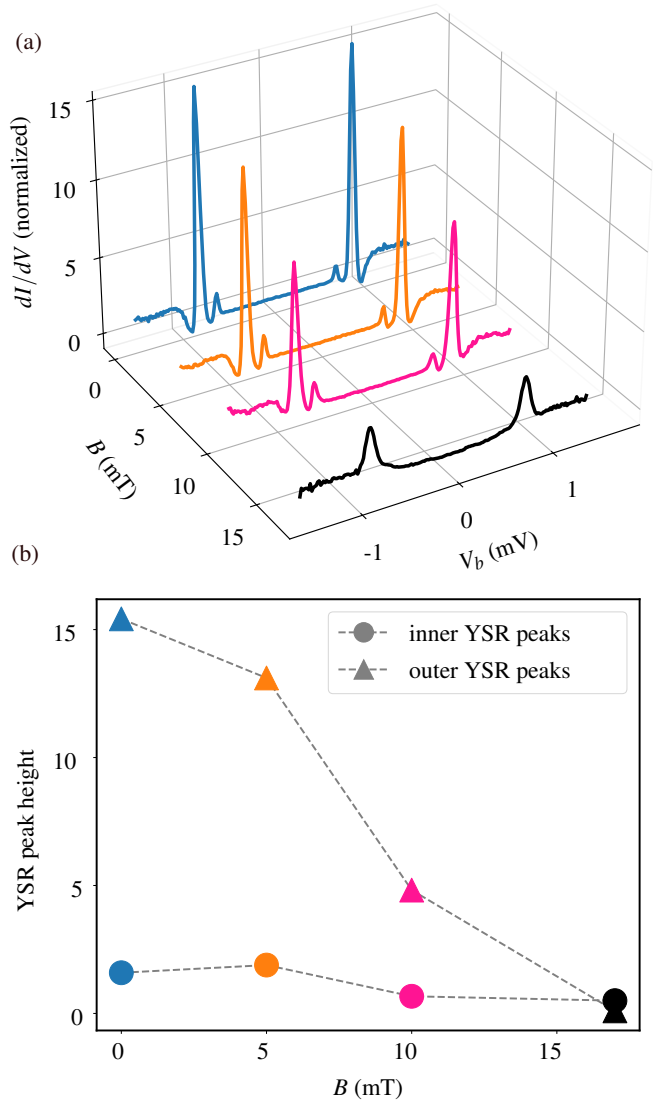
    \caption{ Magnetic field dependence of an effective $S = 3/2$ YSR state embedded in suboxide regions of $\alpha(111)$ tantalum. \textbf{(a)} Evolution of $dI/dV$ spectra of the YSR states in response to out-of-plane magnetic fields, leading to the suppression of the YSR states and the resurgence of the superconducting order parameter at approximately 18 mT. The dashed vertical lines on the side wall represent the projection of the average YSR peak heights. \textbf{(b)} Extracted peak values of the $dI/dV$ spectra of the outer (larger) and inner (smaller) YSR states. Measurements were obtained at a fixed tip position under the following conditions: bias voltage $V_{b} = 1.5$ mV, current setpoint $I_{p}= 100$ pA, lock-in modulation voltage $V_{\text{mod}} = 20$ $\mu$V  and temperature $T= 45$ mK, utilizing a superconducting tip with $\Delta_{\text{tip}} =   208.1$ $\mu$eV.}
\label{fig:Figure3}
\end{figure} 

Here, a crucial point in understanding the suppression of YSR states by a moderate out-of-plane field $(B/B_{c} \sim 0.17)$ is the smearing of the superconducting gap edge. As originally discussed by Shiba\cite{Shiba1968}, a bound state forms within the gap largely because the quasiparticle density of states is sharply peaked at $\Delta$. However, once a field on the order of 15–20 mT is applied, the resulting screening currents and slight Zeeman splitting broaden the coherence peaks and reduce their amplitude, effectively blurring the sharp gap edge. This blurring allows YSR states near the gap to overlap with continuum states above $\Delta$, causing them to become overdamped and eventually disappear. 

The spatial distribution of the YSR states is entirely random, with many having migrated below the surface due to collisions with argon atoms. Detailed mapping and orbital characterization of these states are typically carried out on single-crystal superconducting substrates with atomic resolution and controlled defect creation methods \cite{Shuai-Hua2008, Ruby2015, wang2024}, a level of precision that is unattainable for superconducting films possessing granular surfaces and amorphous oxide layers as used in applications for superconducting electronics.

In an attempt to eliminate regions containing YSR states, likely originating from Ar$+$ milling defects, we extended the milling process for an additional 30 minutes. Surprisingly, this resulted in the observation of numerous regions exhibiting YSR states with a double-pair structure. The inner (smaller) and outer (larger) pairs exhibited binding energies of $410.2$ $\mu$eV and $598.4$ $\mu$eV, respectively, suggesting the presence of localized spins with higher angular momentum (Figure \ref{fig:Figure3}a). 

The observed increase in the number of YSR peaks and their shift toward the Fermi level with extended argon milling can be explained by the following mechanism. During mild argon milling, tantalum–oxygen bonds break randomly, reducing the oxidation state of the central tantalum atoms. These unsatisfied bonds create localized charge and spin centers that remain sufficiently decoupled from the Cooper pairs on the superconducting surface by the remaining oxide layer, resulting in only a negligible exchange interaction. Consequently, the YSR peaks appear asymptotically close to the superconducting coherence peaks. As the milling time increases, more tantalum–oxygen bonds are broken, further reducing the oxidation state of the tantalum atoms. At the same time, the oxide layer is either removed or pushed beneath the surface by argon collisions, bringing these unsatisfied tantalum atoms closer to the superconducting interface. This closer proximity enhances the exchange interaction, and shifts the YSR peaks towards Fermi level.

Under an out-of-plane magnetic field, the behavior of YSR peaks near the coherence peak follows a similar trend as before. These peaks were rapidly suppressed and pushed outside the superconducting gap at approximately 18 mT, while the smaller YSR peaks situated well within the gap gradually broadened and diminished in intensity (Figure \ref{fig:Figure3}b). This behavior is consistent with characteristics of YSR states that extend over a large area and exhibit significant magnetic anisotropy, as discussed in detail in Ref. \cite{Machida2022}. It is noteworthy that magnetism induced by oxygen vacancies at metallic and superconducting interfaces has been widely reported for both strong and weak exchange interaction regimes \cite{Yang2020, yeh2020, Tamir2022, Pritchard2024}. 

In conclusion, our study reveals the unintended consequences of commonly employed cleaning procedures, such as oxygen descumming and argon milling, in the fabrication of tantalum-based superconducting qubits. Although these techniques are critical for preparing the surface and ensuring high-quality Josephson junctions, they also introduce oxygen defects that give rise to magnetic impurities. These impurities, in turn, host YSR bound states within the superconducting gap, which degrade the coherence and performance of tantalum qubits. Our observations highlight the necessity of revisiting and refining the cleaning protocols in qubit fabrication. A delicate balance must be achieved between the removal of oxide layers and the preservation of the pristine superconducting properties of tantalum thin films. An alternative approach could be the use of noble metal passivation layers to prevent the formation of native tantalum oxides in the first place \cite{Chang2024}. 

Refer to the supplementary material for a detailed characterization of the SIS tunnel junction under an applied magnetic field, as well as the extraction of the tantalum superconducting gap and its critical magnetic field.

\begin{acknowledgments}
This work was supported by the German Ministry of Education and Research (BMBF) within the project QSolid (FKZ:13N16151).
\end{acknowledgments}

\section*{author declarations}
\subsection*{Conflict of Interest}
The authors declare that there are no conflicts of interest.

\section*{Author Contributions}

\textbf{Soroush Arabi}:  Investigation (lead); Conceptualization (equal); Data curation (lead); Formal analysis (lead); Methodology (lead); Validation (lead); Visualization (lead); Project administration (lead); Writing - original draft (lead); Writing -  review \& editing (lead). \textbf{Qili Li}: Investigation (equal); Writing – review \& editing (equal). \textbf{Ritika Dhundhwal}: Investigation (equal); Data curation (equal); Writing – review \& editing (equal). \textbf{Dirc Fuchs}: Investigation (equal). \textbf{Thomas Reisinger}: Conceptualization (equal);  Methodology (equal); Writing – review \& editing (equal). \textbf{Ioan Pop}: Conceptualization (equal); Methodology (equal); Writing – review \& editing (equal); Resources (equal); Funding acquisition (equal); Supervision (equal). \textbf{Wulf Wulfhekel}: Conceptualization (equal); Methodology (equal); Writing – review \& editing (equal); Resources (equal); Funding acquisition (equal); Supervision (equal)

\section*{DATA AVAILABILITY}
Data supporting the findings of this study are available from the corresponding author upon reasonable request.

\bibliographystyle{apsrev4-1} 
\bibliography{main}

\end{document}


\maketitle

\begin{abstract}
In this supplementary material, we provide detailed information on the characterization of the superconducting tunnel junction used in the STM measurements to determine the superconducting gap and critical magnetic field of the $\alpha(111)$ tantalum thin film, as reported in the main text.
\end{abstract}

\setcounter{figure}{0} 
\renewcommand{\thefigure}{S\arabic{figure}} 
\renewcommand{\figurename}{FIG.} 

\begin{figure}[H]
    \centering
    \def\svgwidth{1\textwidth} 
    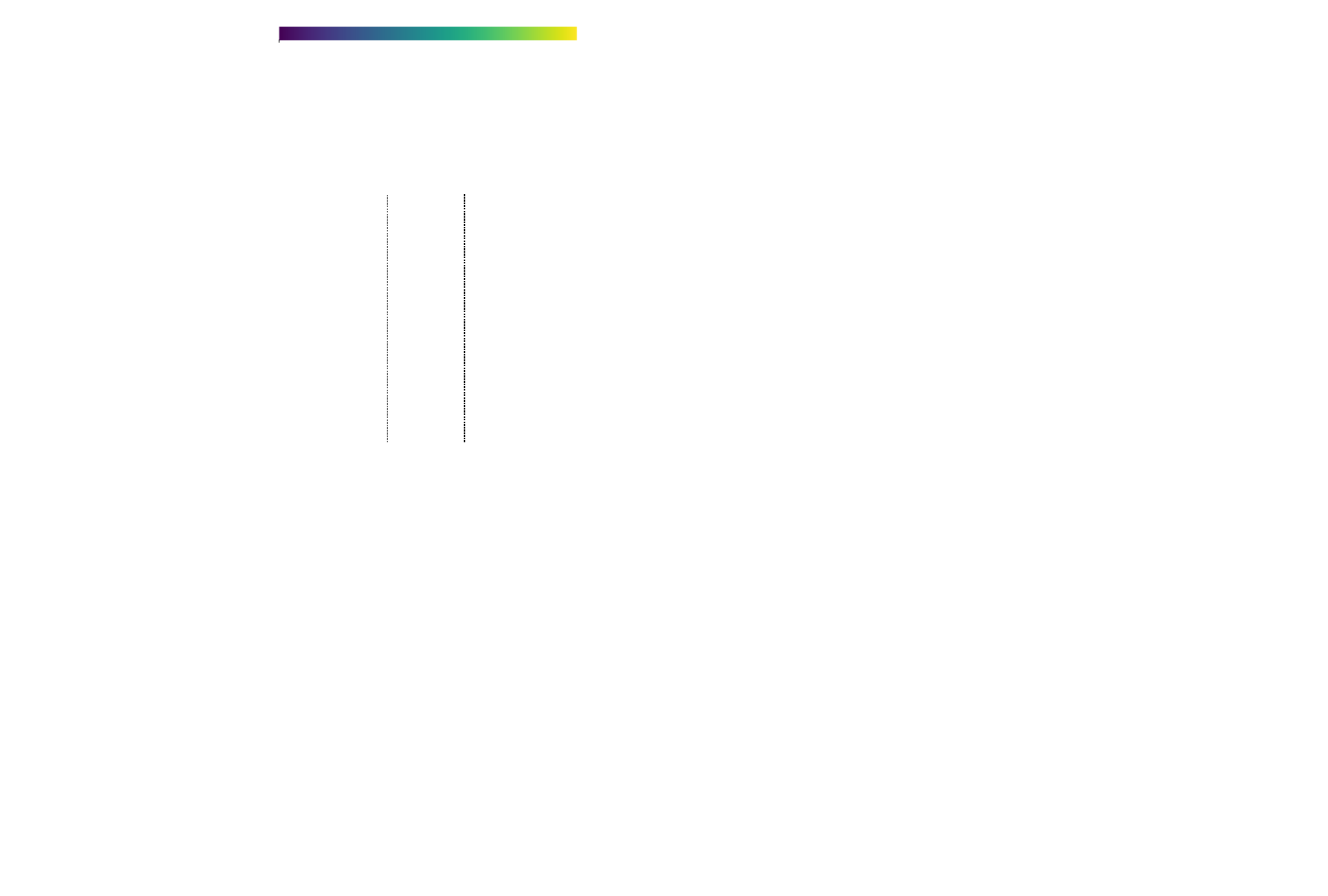
    \caption{Characterization of the SIS tunnel junction under an applied magnetic field. \textbf{(a)} Evolution of the $dI/dV$ spectra of the SIS tunnel junction as a function of out-of-plane magnetic fields, revealing the superconducting gaps of both the sample and the tip. \textbf{(b)} Superconducting gap of the tantalum thin film as a function of magnetic field. \textbf{(c)} Superconducting gap of the tantalum tip as a function of magnetic field. The superconducting gaps for both the sample and the tip were extracted from each $dI/dV$ spectrum using the Dynes formula. The measurements in \textbf{(b)} and \textbf{(c)} were fitted using the Landau-Ginzburg model, $\Delta/\Delta_{0} =  1 - (B/B_{c})^{2}$, yielding zero-field superconducting gaps of $\Delta_{\text{sample,0}} = 600.3\,\mu eV$ and $\Delta_{\text{tip,0}} = 208.1\,\mu eV$, with critical fields of $B_{\text{sample,c}} = 103.1\,\text{mT}$ and $B_{\text{tip,c}} = 1.59\,\text{T}$, respectively. The $dI/dV$ measurements were taken at a fixed tip position under the following conditions: bias voltage $V_{b} = 1.5\,\text{mV}$, current setpoint $I_{p} = 100\,\text{pA}$, lock-in modulation voltage $V_{\text{mod}} = 20\,\mu V$, and temperature $T = 45\,\text{mK}$.}
    \label{fig:supplement}
\end{figure}

\section*{S1. Characterization of the SIS Tunnel Junction}

We extract the superconducting gap and critical magnetic field of $\alpha(111)$ Ta based on the method explained in Figure~\ref{fig:supplement}. We performed $dI/dV$ measurements in an out-of-plane magnetic field. The $dI/dV$ spectrum evolved from a double-gap structure at low magnetic fields for the combined gaps (Figure~\ref{fig:supplement}a, top panel), to a conventional BCS-like single-gap spectrum at higher magnetic fields, representing the tip's superconducting gap (Figure~\ref{fig:supplement}a, bottom panel).

The superconducting gap at each magnetic field was determined by fitting the spectrum with the Dynes function \citep{Dynes1978}, after which the measured gaps as a function of magnetic field were fit using the Ginzburg-Landau model\cite{Casella1964,Meservey1964} (Figure~\ref{fig:supplement}b and S1c). This analysis yielded zero-field superconducting gaps of $\Delta_{\text{sample,0}} = 600.3\,\mu eV$ and $\Delta_{\text{tip,0}} = 208.1\,\mu eV$, with corresponding critical fields of $B_{\text{sample,c}} = 101.7\,\text{mT}$ and $B_{\text{tip,c}} = 1.51\,\text{T}$. Notably, the critical field of the sharp superconducting tip is much larger than that of the sample, owing to the much smaller size of the tip’s apex and the correspondingly smaller magnetic flux passing through it. 

\bibliographystyle{apsrev4-1} 
\bibliography{Supplement} 